\documentclass{aa}  

\usepackage{graphicx}
\usepackage{txfonts}
\usepackage{multirow}
\usepackage{color}

\newcommand{\flux}{egs~s$^{-1}$~cm$^{-2}$}
\newcommand{\msun}{M$_{\odot}$}
\newcommand{\mpy}{\msun~yr$^{-1}$}

\newcommand{\sfr}{$\text{SFR} = 77^{+40}_{-25} $~\mpy} 
\newcommand{\mstar}{$M_{\star} = 8\times 10^9$~\msun} 
\newcommand{\sigSFR}{$\Sigma_{\rm SFR} = 1.6$~\msun~kpc$^{-2}$}
\newcommand{\rhalfFeiiarcsec}{$0.49\pm0.05$}
\newcommand{\rhalfContarcsec}{$0.28\pm0.02$}
\newcommand{\rhalfOiiarcsec}{$0.33\pm0.02$}
\newcommand{\rhalfFeiikpc}{$4.1\pm0.4$}
\newcommand{\rhalfContkpc}{$2.34\pm0.17$}
\newcommand{\rhalfOiikpc}{$2.76\pm0.17$}
\newcommand{\redshift}{$1.29018 \pm 0.00006$}

\newcommand{\lya}{\hbox{{\rm Ly}\kern 0.1em$\alpha$}}
\newcommand{\Ly}{\hbox{{\rm Ly}\kern 0.1em$\alpha$}}
\newcommand{\Ha}{\hbox{{\rm H}\kern 0.1em$\alpha$}}
\newcommand{\Hb}{\hbox{{\rm H}\kern 0.1em$\beta$}}
\newcommand{\OII}{[\hbox{{\rm O}\kern 0.1em{\sc ii}}]}
\newcommand{\OIII}{[\hbox{{\rm O}\kern 0.1em{\sc iii}}]}
\newcommand{\FeII}{\hbox{{\rm Fe}\kern 0.1em{\sc ii}}}
\newcommand{\FeIIs}{\ion{Fe}{II}*}

\begin{document} 

	\title{Galactic Winds with MUSE: A Direct Detection of \ion{Fe}{II}* Emission from a $z=1.29$ Galaxy  \thanks{ Based on observations of the {\it Hubble} Deep Field South made with ESO telescopes at the
La Silla Paranal Observatory under program ID 60.A-9100(C). Advanced data products are available at http://muse-vlt.eu/ science.}   }

   \author{Hayley Finley\inst{1,2}
          \and
          Nicolas Bouch\'e\inst{3}
          \and
          Thierry Contini\inst{1,2}
          \and
          Beno\^it Epinat \inst{1,2,4}
\and Roland Bacon \inst{5}
\and Jarle Brinchmann \inst{6,7}
\and Sebastiano Cantalupo \inst{8}
\and Santiago Erroz-Ferrer \inst{8}
\and Raffaella Anna Marino \inst{8}
\and Michael Maseda \inst{6}
\and Johan Richard \inst{5}
\and Anne Verhamme \inst{5,9}
\and Peter M. Weilbacher \inst{10}
\and Martin Wendt \inst{10,11}
\and Lutz Wisotzki \inst{10}
          }

   \institute{Universit\'e de Toulouse, UPS-OMP, 31400 Toulouse, France 
   \email{hayley.finley@irap.omp.eu}
   \and
   IRAP, Institut de Recherche en Astrophysique et Plan\'etologie, CNRS, 14 avenue \'Edouard Belin, 31400 Toulouse, France 
   \and 
   IRAP, Institut de Recherche en Astrophysique et Plan\'etologie, CNRS, 9 avenue Colonel Roche, 31400 Toulouse, France 
   \and
   Aix Marseille Univ, CNRS, LAM, Laboratoire d'Astrophysique de Marseille, Marseille, France
   \and
  CRAL, Observatoire de Lyon, CNRS, Université Lyon 1, 9 Avenue Ch. André, F-69561 Saint Genis Laval Cedex, France 
  \and
  Leiden Observatory, Leiden University, P.O. Box 9513, 2300 RA Leiden, The Netherlands 
  \and
  Instituto de Astrof{\'i}sica e Ci{\^e}ncias do Espa{\c{c}}o, Universidade do Porto, CAUP, Rua das Estrelas, PT4150-762 Porto, Portugal 
  \and
  ETH Zurich, Institute of Astronomy, Wolfgang-Pauli-Str. 27, CH-8093 Z\"urich, Switzerland 
 \and 
 Observatoire de Genève, Université de Genève, 51 Ch. des Maillettes, 1290 Versoix, Switzerland
 \and 
 Leibniz-Institut für Astrophysik Potsdam (AIP), An der Sternwarte 16, D-14482 Potsdam, Germany 
 \and 
 Institut für Physik und Astronomie, Universität Potsdam,Karl-Liebknecht-Str. 24/25, 14476 Golm, Germany 
   }
          
   \date{}

 
  \abstract{
Emission signatures from galactic winds provide an opportunity to directly map the outflowing gas, but this is traditionally challenging because of the low surface brightness. 
Using very deep observations  (27 hours) of the {\it Hubble} Deep Field South with the Multi Unit Spectroscopic Explorer (MUSE) instrument, we identify signatures of an outflow in both emission and absorption from a spatially resolved galaxy at $z=1.29$ with a stellar mass \mstar, star formation rate \sfr, and star formation rate surface brightness \sigSFR\ within the \OII\,$\lambda\lambda3727,3729$ half-light radius $R_{1/2,\,\OII}=$\rhalfOiikpc~kpc. 
From a component of the strong resonant \ion{Mg}{II} and \ion{Fe}{II} absorptions at $-350$~km~s$^{-1}$, we infer a mass outflow rate that is comparable to the star formation rate. 
We detect non-resonant \ion{Fe}{II}* emission, at $\lambda$2626, $\lambda$2612, $\lambda$2396, and $\lambda$2365, at $1.2 - 2.4 - 1.5 - 2.7 \times 10^{-18}$ \flux\ respectively.
These flux ratios are consistent with the expectations for optically thick gas. By combining the four non-resonant \FeIIs\ emission lines, we spatially map the \ion{Fe}{II}* emission from an individual galaxy for the first time.
The \FeIIs\ emission has an elliptical morphology that is roughly aligned with the galaxy minor kinematic axis, and its integrated half-light radius, $R_{1/2,\,\FeIIs}=$~\rhalfFeiikpc~kpc, is 50\%\ larger than the stellar continuum ($R_{1/2,\star}\simeq$~\rhalfContkpc) or the \OII\ nebular line. Moreover, the \ion{Fe}{II}* emission shows a blue wing extending up to $-400$ km s$^{-1}$, which is more pronounced along the galaxy minor kinematic axis and reveals a C-shaped pattern in a $p-v$ diagram along that axis. These features are consistent with a bi-conical outflow. 
}


   \maketitle	
%


\section{Introduction}

Galactic winds, driven by the collective effect of hot stars and supernovae explosions, play a major role in regulating galaxy evolution.
By expelling enriched matter beyond the halo, galactic winds can address discrepancies between observations and $\Lambda$CDM models that over-predict the number of low-mass galaxies \citep{2012RAA....12..917S} and enrich the intergalactic medium \citep{2008MNRAS.387..577O, 2016MNRAS.459.1745F}.
Likewise, galactic winds may play a major role in regulating the mass-metallicity relation \citep{2008MNRAS.385.2181F, 2013ApJ...772..119L,TremontiC_04a}. Therefore, quantifying the mass fluxes of galactic outflows (and their extents) is  necessary to gain a complete understanding of galaxy evolution.

However, while  galactic winds appear ubiquitous \citep[e.g.,][]{2005ARA&A..43..769V, 2009ApJ...692..187W, SteidelC_10a, 2010ApJ...719.1503R, 2014ApJ...794..156R,  2012ApJ...760..127M, 2015ApJ...809..147H, 2015ApJ...815...48Z, 2015ApJ...811..149C},
observational constraints for the physical properties of galactic outflows, including their extents and mass outflow rates, are sparse. 
Traditional "down the barrel" 1D galaxy spectroscopy provides direct constraints on the wind speed from the blue-shifted absorption lines but cannot constrain the physical extent of outflows, leading to large uncertainties in outflow rates.
Techniques that use a background source  can address this question.

For instance, the background quasar technique provides constraints on the physical extent of gas flows from the impact parameter between the galaxy and the absorbing gas \citep[e.g.][]{2012MNRAS.426..801B, 2012ApJ...760L...7K, 2015ApJ...804...83S, 2016arXiv160503412S, 2016MNRAS.457..903P, 2016ApJ...820..121B, 2016MNRAS.458.3760S}. These recent studies have made progress investigating the kinematics, orientation, and extent of gas flows around star forming  galaxies.
As a variation on this technique, spectroscopy against a background galaxy probes absorption from the foreground galaxy halo over a larger solid angle \citep[e.g.,][]{2005ApJ...629..636A, 2010ApJ...712..574R,SteidelC_10a, 2011ApJ...743...10B, 2014ApJ...784..108B, 2016ApJ...824...24D}.
However, these constraints on the physical extent of outflows are usually limited due to their one-dimensional nature, except for \citet{CazzoliG_16a}. 
Mapping the extent of gas flows in two dimensions is critical to better constrain mass outflow rates.

Mapping outflows in emission, such as for M82 \citep[e.g.,][]{1998ApJ...493..129S, 1999ApJ...523..575L} and other nearby galaxies \citep[e.g.,][]{HeckmanT_95a,Cecil_01a,VeilleuxS_02a,Matsubayashi_09a,Moiseev_10a}, is difficult at high redshift, because the emitting gas inherently has a very low surface brightness.
Beyond the local universe, galaxies with emission signatures from outflows are beginning to be detected.
Currently, rest-frame UV and optical spectroscopy use three types of emission signatures to map the extent of outflows:   
the nebular, resonant, and non-resonant emission lines. 
The most common nebular emission lines seen in \ion{H}{II} regions are hydrogen recombination and forbidden lines, such as [\ion{O}{II}]\,$\lambda\lambda$3727,3729.
A transition is resonant when a photon can be absorbed from the ground state and re-emitted to the same lowest level of the ground state, as for Lyman-alpha and the \ion{Mg}{II} $\lambda\lambda2796,2803$ transitions.
A transition is non-resonant when the photon can be re-emitted to an excited level of a ground state that has multiple levels due to fine structure splitting.
Non-resonant transitions are commonly denoted with a *, like \ion{Fe}{II}*. 
Due to the slight energy difference between the ground and excited states,  photons from non-resonant emission no longer have the correct wavelength to be re-absorbed through a resonant transition and instead escape.
In other words, the gas is optically thin to photons that are emitted through a non-resonant transition.

The first type of emission signature (nebular lines) from outflows can appear as a broad component in nebular emission lines such as \Ha. 
Such broad component is regularly seen in local  Ultra-Luminous Infra-Red Galaxies  \cite[ULIRGs, e.g.][]{SotoK_12a,ArribasS_14a,Garcia_15a} and more recently in normal star-forming galaxies \citep{WoodC_15a,CiconeC_16a}.
At high redshifts, \cite{2012ApJ...761...43N} detected a broad \Ha\ component in composite spectra of $z \sim 2$ star-forming galaxies and \cite{2011ApJ...733..101G} observed this broad component in a few individual galaxies. \cite{2012ApJ...761...43N} found that the broad emission is spatially extended beyond the half-light radius, $R_{1/2}$.

The second possible emission signature of outflows comes from resonant transitions such as \lya, a line which is often more extended than the stellar continuum \citep[e.g.][]{SteidelC_11a,Matsuda_12a,WisotzkiL_16a} but might be strongly affected by dust absorption. 
Emission from resonant metal lines, such as \ion{Si}{ii}, \ion{Fe}{ii}, or \ion{Mg}{ii}, is less affected by dust and may be observed as P-cygni profiles.
The relative strength between the (mostly) blueshifted absorption and (mostly) redshifted emission dictates whether the signature appears as a traditional P-cygni profile or as emission `infilling'. Contrary to the resonant \ion{Fe}{II} lines observed across a similar wavelength range (\ion{Fe}{II}\,$\lambda2344$, $\lambda\lambda 2374,2382$, and $\lambda\lambda 2586,2600$), the \ion{Mg}{II} doublet is particularly sensitive to emission infilling, since its lower energy level does not have
fine structure splitting. As a result of the different possible relative strengths of the emission and absorption components, observed profiles for the resonant \ion{Mg}{II}\,$\lambda\lambda 2796,2803$ transitions vary greatly for different star-forming galaxies \citep{2009ApJ...692..187W, 2011ApJ...728...55R, 2011ApJ...743...46C, 2012ApJ...759...26E,  2012ApJ...760..127M,  2013ApJ...770...41M, 2013ApJ...774...50K}.

The third possible signature of outflows in emission is from non-resonant transitions such as \ion{C}{II}*, \ion{Si}{II}*
\citep[e.g.,][]{ShapleyA_03a}
or \ion{Fe}{II}* \citep[e.g.,][]{2011ApJ...728...55R}.
Detecting non-resonant emission  typically requires stacking  hundreds of galaxy spectra. Using more than 800 Lyman break galaxies (LBGs) at $z > 2$, \cite{ShapleyA_03a} first detected \ion{Si}{II}* in the composite spectrum, and \cite{2012ApJ...749....4B} more recently detected \ion{C}{II}* and \ion{Si}{II}* in the composite spectrum of 59 LBGs.
Since the non-resonant \ion{Fe}{II}* lines are at redder wavelengths than \ion{C}{II}* and \ion{Si}{II}*, they are practical for investigating outflows at lower redshifts, like $z\sim1$. 
Based on comparing composite spectra from samples of $\sim 100$ or more star-forming galaxies at $z \sim 1-2$ \citep{2012ApJ...759...26E, 2013ApJ...774...50K, 2014ApJ...793...92T},
\ion{Fe}{II}* emission may vary with galaxy properties, such as galaxy mass and dust attenuation.
\cite{2011ApJ...743...46C} present individual spectra with different combinations of blue-shifted absorption, resonant \ion{Mg}{II} emission, and non-resonant \ion{Fe}{II}*.
In two notable direct detections of \ion{Fe}{II}* emission from galaxies at $z = 0.694$ and $z = 0.9392$ \citep{2011ApJ...728...55R, 2012ApJ...760..127M}, the non-resonant emission is observed along with blue-shifted absorption lines and resonant \ion{Mg}{II} emission, allowing the authors to constrain and model the outflows.
Similarly, \cite{2014ApJ...791L..19J} use non-resonant \ion{C}{II}* and \ion{Si}{II}* emission in UV spectra of four green pea galaxies at $z \sim 0.14 - 0.2$ to infer the geometry of their outflows. 

These studies provide information about outflow properties on galactic scales, but it is also possible to characterize outflows from individual star-forming regions across $z > 1$ galaxies thanks to adaptive optics or gravitational lensing \citep[i.e.,][Patricio et al.\ in prep.]{2011ApJ...733..101G, 2014ApJ...790...44R, 2016A&A...585A..27K, 2016MNRAS.458.1891B}.
Using adaptive optics, \cite{2011ApJ...733..101G} identify star-forming regions in five $z > 2$ galaxies and argue that bright regions (or clumps) with a broad component in the nebular emission are the launch sites for massive galactic winds.
With the benefit of gravitational lensing, \cite{2016A&A...585A..27K} characterize \ion{Mg}{II} emission, \ion{Fe}{II}*\,$\lambda\lambda2612, 2626$ emission, and \ion{Fe}{II} absorption from multiple star-forming regions across a supernova host galaxy at $z = 1.49$ at locations both associated with and independent of the supernovae explosion.
\cite{2016MNRAS.458.1891B} likewise detect blueshifted \ion{Fe}{II} and \ion{Mg}{II} absorptions, redshifted \ion{Mg}{II} emission, and non-resonant \ion{Fe}{II}*\,$\lambda\lambda2612, 2626$ emission in four star-forming regions  of a gravitationally lensed galaxy at $z = 1.70$, but find that the outflow properties vary from region to region. 
Spatially resolved observations suggest that outflow properties could be localized and strongly influenced by the nearest star-forming clump. 

Despite advances from these diverse studies, we have not yet been able to map the morphology and extent of outflows from individual galaxies beyond the local universe. 
The new generation of integral field spectrographs, the Multi Unit Spectroscopic Explorer \citep[MUSE;][]{2015A&A...575A..75B} on the VLT and the Keck Cosmic Web Imager \citep{2012SPIE.8446E..13M}, are well-suited for studying galactic winds in emission and tackling this challenge. 
While slit spectroscopy can inadvertently miss scattered emission if the aperture does not cover the full extent of the outflowing envelope \citep{2015ApJ...801...43S}, integral field observations eliminate aperture effects for distant galaxies, making emission signatures easier to detect. 
The combined spatial and spectroscopic data facilitate characterizing the morphology and kinematics of both star-forming galaxies and the outflows they produce.

In this paper, we analyze galactic wind signatures from a spatially resolved star-forming galaxy at $z = 1.2902$ observed with MUSE. We present the observations in Section~\ref{sect:data} and summarize the galaxy properties in Section~\ref{sect:galProp}.  
With the integrated 1D MUSE galaxy spectrum, we characterize outflow signatures from \ion{Fe}{II}, \ion{Mg}{II}, and \ion{Mg}{I} transitions in absorption and \ion{Fe}{II}* transitions in emission in Section~\ref{sect:1Dspec}.
We then investigate the spatial extent and the kinematic properties of the \ion{Fe}{II}* emission in Sections~\ref{section:2D} and \ref{section:kinematics}, respectively. 
In Section~\ref{sect:discuss}, we compare our observations with radiative transfer wind models and estimate the mass outflow rate.
We review our findings in Section~\ref{sect:conclusions}.
Throughout the paper, we assume a $\Lambda$CDM cosmology with $\Omega_{\rm m} = 0.3$, $\Omega_{\Lambda} = 0.7$, and $H_0 = 70$~km~s$^{-1}$~Mpc$^{-1}$. With this cosmology, 1 arcsec corresponds to 8.37~kpc at the redshift of the galaxy.


\section{Data}
\label{sect:data}

MUSE fully covers the wavelength range $4650-9300\ \AA$ with 1.25 \AA\ per spectral pixel. The field of view spans $1\arcmin \times 1\arcmin$ with a pixel size of 0.2\arcsec. The instrument is notable both for its high throughput, which reaches 35\% at 7000 \AA\ (end-to-end including the telescope), and its excellent image quality sampled at 0.2\arcsec\ per spaxel. 
While MUSE opens new avenues to address a wide variety of scientific questions, these two characteristics  make the instrument optimal for deep field observations.

As part of commissioning data taken during July and August 2014, MUSE observed a $1\arcmin \times 1\arcmin$ field of view in the Hubble Deep Field South (HDFS) for a total integration time of 27 hours. The final data cube is a 5$\sigma$-clipped mean of 54 individual exposures that were taken in dark time under good seeing conditions ($0.5\arcsec-0.9\arcsec$). The 1$\sigma$ emission-line surface brightness limit for this cube is $1 \times 10^{-19}\text{ erg\ s}^{-1}\ \text{cm}^{-2}\ \text{arcsec}^{-2}$. 
The MUSE observations provided spectroscopic redshifts for 189 sources with magnitude $I_{814} \leq 29.5$ (8 stars and 181 galaxies), a factor-of-ten increase over the 18 previously-measured spectroscopic redshifts in this field. A catalogue of sources in the MUSE HDFS field includes the redshifts, emission-line fluxes, and 1D spectra. The observations, the data cube, and an overview of scientific exploitations are fully described in \cite{2015A&A...575A..75B}. Both the data cube and the catalogue of sources are publicly available.\footnote{\url{http://muse-vlt.eu/science/hdfs-v1-0/}}

The deep IFU observations reveal emission from \ion{Fe}{II}* transitions directly detected from one galaxy in the MUSE HDFS. 
The galaxy has ID \#13 in the MUSE catalogue, with coordinates $\alpha = 22\text{h}\,32\text{m}\,52.16\text{s}$, $\delta = -60^{\circ}\,33\arcmin\,23.92\arcsec$ (J2000) and magnitude $I_{814} = 22.83 \pm 0.005$. It is part of a 9-member group at $z \simeq 1.284$, discussed in \cite{2015A&A...575A..75B}, that also includes two AGN and an interacting system with tidal tails.
This direct detection of a galaxy with \ion{Fe}{II}* emission offers a new opportunity to characterize galactic winds.

\section{Galaxy Properties}
\label{sect:galProp}

\begin{table} 
\caption{Galaxy ID\#13 properties from \citet{2016A&A...591A..49C}}  
\label{table:galProps}
\centering                                      
\begin{tabular}{l c }          
\hline\hline                       
    Morphological Analysis                     & HST $+$ {\sc Galfit}  \\
\hline             
    Position angle ($^{\circ}$)                & $-45.9 \pm 1.9$       \\ 
    Inclination $i$ ($^{\circ}$)               & $33 \pm 5$            \\ 
    Half-light radius (kpc)					   & $2.1\pm 0.03$       \\
\hline\hline
    Kinematic Analysis                         & MUSE 2D/3D            \\   
\hline
    Position angle ($^{\circ}$)                &  $-14$/$-13$           \\  
	Inclination $i$ ($^{\circ}$)               &   $+$28/$+17$           \\      
    Max.\,rotational velocity (km s$^{-1})$    &  $+24$/$+44$           \\
    Velocity dispersion (km s$^{-1})$          &  $+48$/$+46$           \\
\hline\hline
    Photometric Analysis 					   & SED fitting            \\   
\hline
	Visual extinction A$_\text{V}$ (mag)             & $1.20^{+0.59}_{-0.26}$ \\
    $\log\,(M_{\star})$ (M$_{\odot}$)            & $9.89 \pm 0.11$  \\
    $\log\,(\text{SFR})$ (M$_{\odot}$ yr$^{-1}$) & $1.89 \pm 0.18$  \\
\hline                                     
\end{tabular}
\end{table}

\begin{figure*}
\resizebox{\hsize}{!}{\includegraphics{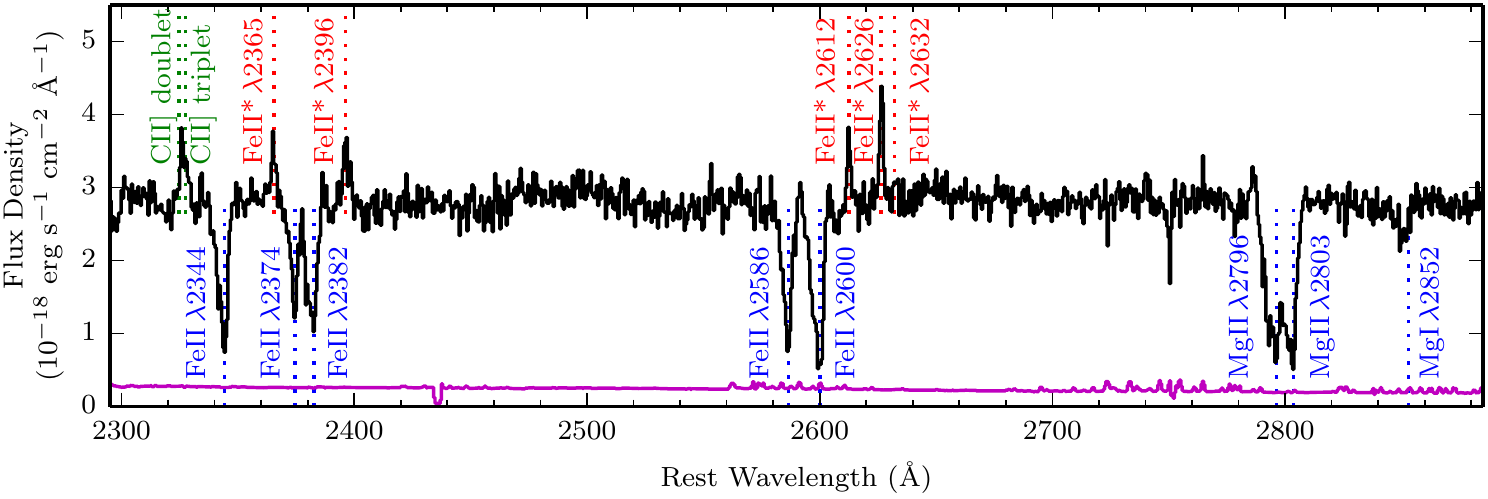}}
\caption{Vacuum rest-frame 1D spectrum of the MUSE HDFS galaxy ID\#13 covering the \ion{Fe}{II} and \ion{Mg}{II} transitions. The spectrum is in black with the $1\sigma$ error in magenta. Resonant transitions detected in absorption are labeled in blue. Non-resonant \ion{Fe}{II}* transitions detected in emission are labeled in red. The \ion{C}{II}] nebular emission, which is a blend of five transitions, is labeled in green. 
}
\label{fig:spec1D}
\end{figure*}

Galaxy ID\#13 is part of a sample of 28 spatially resolved galaxies that \cite{2016A&A...591A..49C} selected from the MUSE HDFS according to the criterion that the brightest emission line covers at least 20 spatial pixels with a signal-to-noise ratio (S/N) higher than 15. For this galaxy, emission from the [\ion{O}{II}]\,$\lambda\lambda3727,3729$ doublet is the dominant feature in the MUSE spectrum.
We determined the galaxy systemic redshift from a $p$-$v$ diagram extracted from the MUSE data cube along the galaxy kinematic major axis by fitting a double Gaussian profile to the [\ion{O}{II}]\,$\lambda\lambda3727,3729$ emission at each position along the slit.
The systemic redshift of $z = 1.29018 \pm 0.00006$ is the mean value between the two asymptotes of the rotation curve.

\cite{2016A&A...591A..49C} investigated the morphological and kinematic properties of the galaxy ID\#13, as part of the MUSE HDFS spatially resolved galaxy sample.
They constrained the morphology from HST images in the F814W band by modelling the galaxy with {\sc Galfit} \citep{2002AJ....124..266P} as a bulge plus an exponential disk.
\cite{2016A&A...591A..49C} then performed the kinematic analysis with two different techniques: a traditional 2D line-fitting method with the Camel algorithm \citep{2012A&A...539A..92E,2016A&A...591A..49C} combined with a 2D rotating disk model, which requires prior knowledge of the galaxy inclination, and a 3D fitting algorithm, GalPaK$^{\text{3D}}$ \citep{2015AJ....150...92B}, which simultaneously fits the morphological and kinematic parameters directly from the MUSE data cube. 
The parameters from the 2D and 3D models are in good agreement overall (see Table~\ref{table:galProps}).

From the morphological analysis on the HST images, galaxy ID\#13 is compact with a disk scale length of $R_d=1.25$ kpc (correspondingly $R_{1/2}=2.1$~kpc) and has a low inclination angle of $i=33^{\circ}$.
The inclination from 3D fitting yields a lower value of $\sim 20^{\circ}$. The disagreement likely arises from an asymmetric morphology seen in the HST images, since statistically the two techniques measure inclinations that are in good agreement \citep{2016A&A...591A..49C}.
The galaxy also shows a misalignment between the morphological position angle measured from the HST image, $-46^{\circ}$, and the MUSE kinematic position angle, $-13^{\circ}$, again likely due to the asymmetric light distribution that only appears at higher spatial resolution. Regardless, the galaxy has a low inclination with $i\sim 20^{\circ} - 30^{\circ}$.

From the kinematic analysis on the MUSE data, the velocity field has a low gradient, $\pm 10$~km~s$^{-1}$, a low maximum velocity, $24$~km~s$^{-1}$,
and a velocity dispersion of $45-50$~km~s$^{-1}.$ Therefore, non-circular motions dominate the gas dynamics within the disk, with $\text{V}/\sigma \approx 0.5$, i.e. below the commonly-used $\text{V}/\sigma \leq 1$ threshold for identifying dispersion-dominated galaxies. Note that the different maximum velocities from the 2D and 3D methods are entirely due to the different inclination values (Table~\ref{table:galProps}). Nonetheless, the ratio remains $\text{V}/\sigma \lesssim 1$ for the range of possible inclinations, $17^{\circ}-33^{\circ}$.

\cite{2016A&A...591A..49C} estimated the visual extinction, $A_V = 1.20$~mag, stellar mass, \mstar, and star formation rate \sfr, from Stellar Population Synthesis using broad-band visible and near infra-red photometry~\footnote{The [OII]-derived SFR for a \cite{2003ApJ...586L.133C} IMF is 65~M$_{\odot}$ yr$^{-1}$ using the \cite{2004AJ....127.2002K} calibration, which also yields an extinction of $A_V=1.5$ in the gas.}.  
The galaxy ID\#13 is one of the most massive of the 28 spatially-resolved galaxies in the MUSE HDFS sample and also has the highest star formation rate (SFR). 
This SFR places galaxy ID\#13 above the main sequence \citep{2007A&A...468...33E, 2011ApJ...730...61K, 2014ApJ...795..104W, 2016ApJ...817..118T} by almost 1 dex, indicating that this galaxy is undergoing a starburst with a high specific SFR of sSFR~$=10$~Gyr$^{-1}$.
The starburst phase of galaxy evolution can produce large-scale outflows when many short-lived massive stars explode as supernovae.

The properties of this galaxy are conducive to detecting signatures from galactic winds.
The low inclination angle favors observing blue-shifted absorptions, given that this signature increases substantially towards face-on galaxies \citep{2010AJ....140..445C, 2012ApJ...758..135K, 2014ApJ...794..156R}.
The [\ion{O}{II}] luminosity ($\sim 10^{43}$~erg~s$^{-1}$) and rest-frame equivalent width ($\sim 50$~\AA, see Table~\ref{table:trans}) indicate that the galaxy ID\#13 is also well-suited for investigating winds in emission, since \ion{Fe}{II}* and \ion{Mg}{II} emission correlate with $L_{\ion{O}{II}}$ or [\ion{O}{II}] rest-frame equivalent width \citep{2013ApJ...774...50K, 2015ApJ...815...48Z}.

\begin{figure}[t!]
\resizebox{\hsize}{!}{\includegraphics{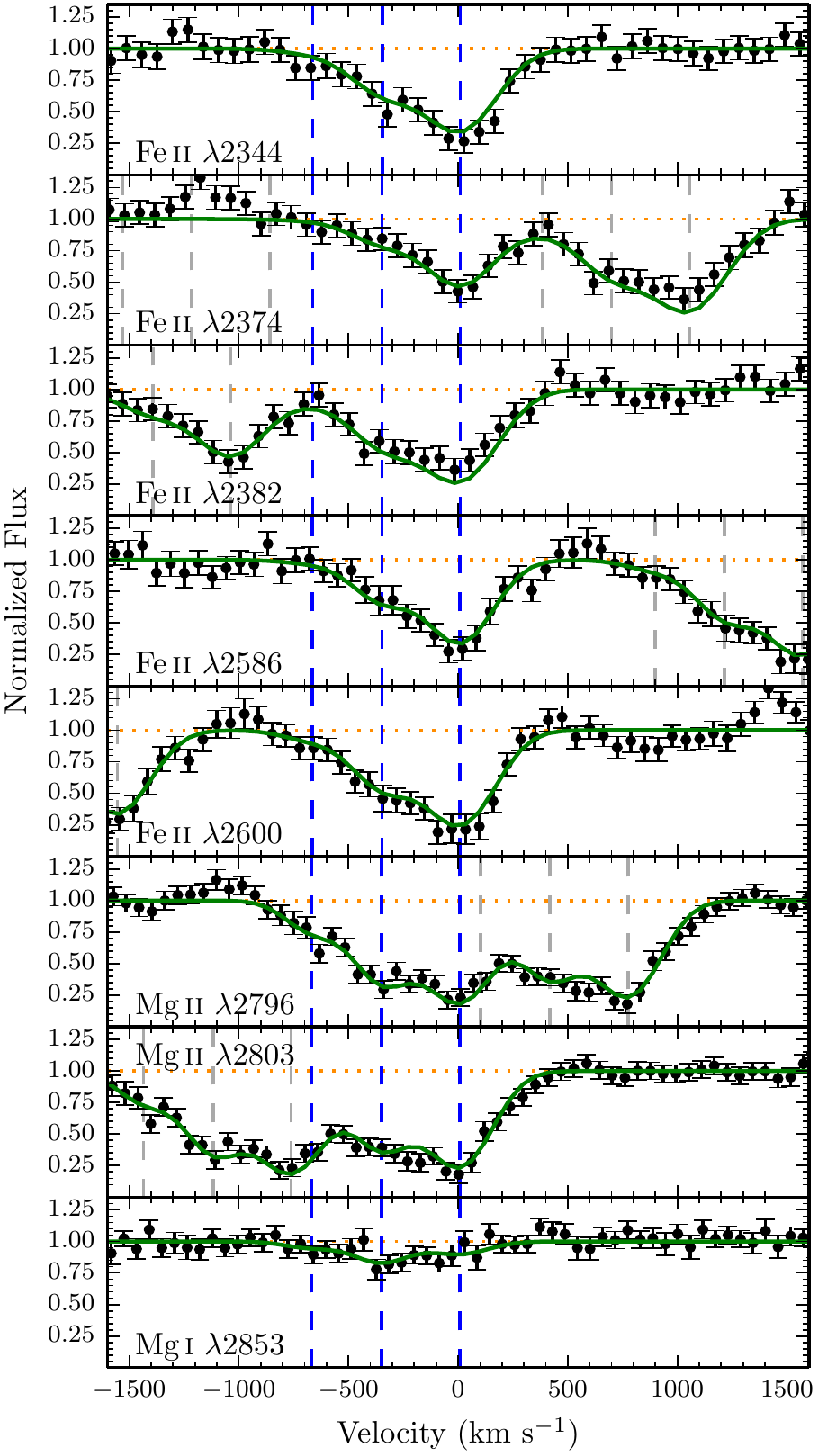}}
\caption{\ion{Fe}{II}, \ion{Mg}{II}, and \ion{Mg}{I} transitions detected in absorption in the 1D MUSE spectrum. Error bars show the 1-$\sigma$ error on the flux (black), and the green curve traces the fit to the absorption profiles. Zero velocity is relative to the galaxy systemic redshift, $z = 1.2902$. Vertical blue dashed lines mark the three components used to fit each absorption, and gray dashed lines show components that are part of neighboring transitions. The asymmetric absorption profiles indicate significant blue-shifted absorption. }
\label{absFits}
\end{figure}

\begin{figure}[t!]
\resizebox{\hsize}{!}{\includegraphics{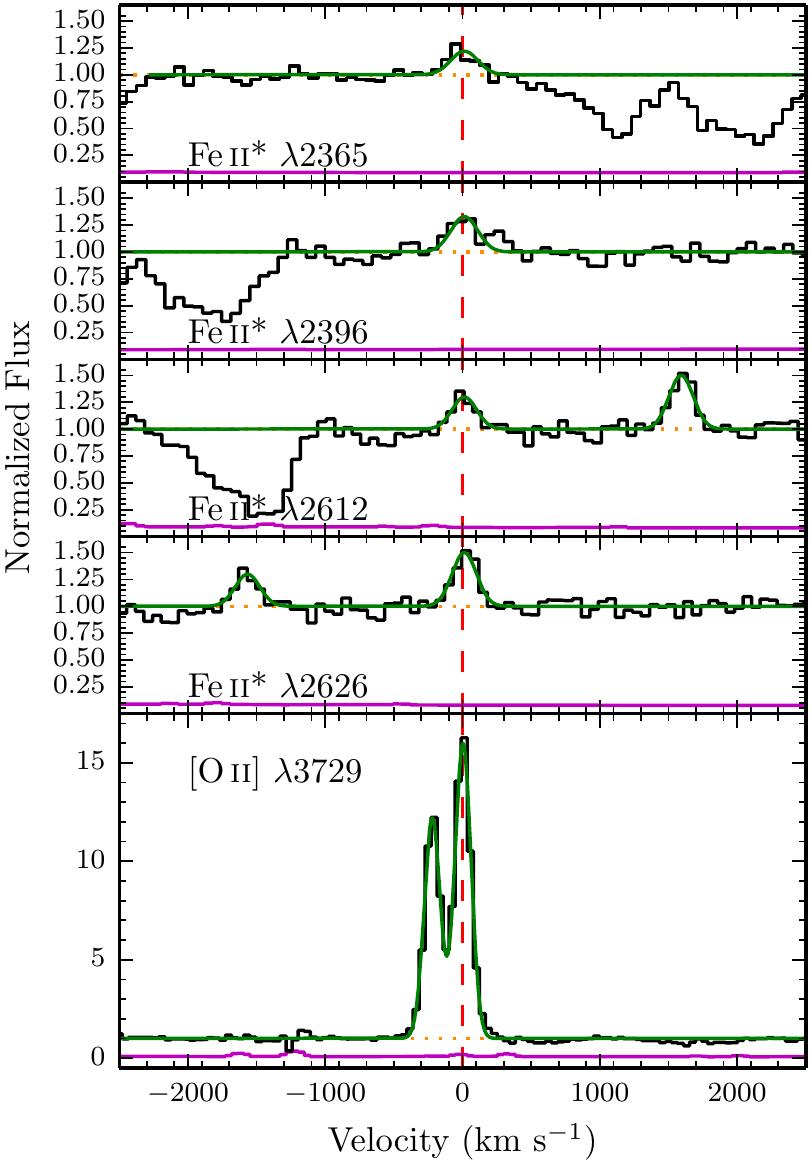}}
\caption{\ion{Fe}{II}* and [\ion{O}{II}] emission peaks detected in the normalized 1D MUSE spectrum. The green curve traces joint Gaussian fits to the four \ion{Fe}{II}* emission peaks and the [\ion{O}{II}] doublet, respectively. Zero velocity, indicated with the vertical red dashed line, is relative to the galaxy systemic redshift, $z = 1.2902$, measured from the [\ion{O}{II}] emission.}
\label{emissFits}
\end{figure}

\begin{table*}[p!]
\caption{Absorption rest-frame equivalent widths for the three sub-components in Figure~\ref{absFits}}  
\label{table:eqW}
\centering                                      
\begin{tabular}{l c c c c c c c c}          
\hline   
\multicolumn{2}{c}{Components}& A & B & C & \multicolumn{2}{c}{Total}{}\\
\multicolumn{2}{c}{Redshift} &  $1.28514$ & $1.28752$ & $1.29024$ & & & \\
\multicolumn{2}{c}{$\Delta v$ (km~s$^{-1}$)}&  $-660 \pm 28$~~ & $-349 \pm 12$~~ & $+8.5 \pm 6.5$ & & & \\
\hline
 Transition  &Multiplet&  W$_{0, \text{fit}}$  & W$_{0, \text{fit}}$ & W$_{0, \text{fit}}$ & W$_{0, \text{fit}}$  & W$_{0, \text{flux}}$  \\ 
  &  &($\AA$) & ($\AA$) & ($\AA$)& ($\AA$)& ($\AA$)   \\
  (1) & (2) & (3) & (4) & (5) & (6) & (7)  \\
\hline   
 \ion{Fe}{II}\,$\lambda2344$ & FeII UV3 & 0.13 & 0.93 & 2.06 &   3.09 &  3.47$\pm$0.24 \\
 \ion{Fe}{II}\,$\lambda2374$ & FeII UV2 &0.04 & 0.55 & 1.70 &   2.42 &  2.20$\pm$0.22 \\
  \ion{Fe}{II}\,$\lambda2382$ & FeII UV2\tablefootmark{a} &  0.24 & 1.17 & 2.36 &   3.69 &  3.27$\pm$0.22  \\
\ion{Fe}{II}\,$\lambda2586$ &FeII UV1 &  0.10 & 0.91 & 2.15 &   3.14 &  3.14$\pm$0.26 \\
\ion{Fe}{II}\,$\lambda2600$ & FeII UV1 & 0.24 & 1.24 & 2.52 &   3.97 &  4.28$\pm$0.25 \\
\hline
\ion{Mg}{II}\,$\lambda2796$ &&  0.98 & 1.64 & 2.82 &   5.51 &  5.09$\pm$0.18 \\
\ion{Mg}{II}\,$\lambda2803$ && 0.93 & 1.89 & 2.63 &   4.47 &  4.90$\pm$0.17 \\
\ion{Mg}{I}\,$\lambda2853$  && 0.16 & 0.49 & 0.31 &   0.94 &  0.86$\pm$0.21 \\
\hline              
\end{tabular}
\tablefoot{Column (1): Absorption line. Column (2): Multiplet associated with transition.
Column (3): Equivalent width for component A.
Column (4): Equivalent width for component B.
Column (5): Equivalent width for component C.
Column (6): Total equivalent width measured from fits.
Column (7): Total equivalent width measured from the spectrum.
}
\tablefoottext{a}{\FeII\,$\lambda$2382 is a pure resonant absorption line with no associated \FeII* emission.}
\end{table*}

\section{Absorption and emission profiles from the 1D spectrum}
\label{sect:1Dspec}

In this section, we analyze the galaxy ID\#13 1D spectrum extracted from the MUSE data using a white-light weighting scheme.
The 1D MUSE spectrum (Figure~\ref{fig:spec1D}) reveals resonant \ion{Fe}{II}, \ion{Mg}{II}, and \ion{Mg}{I} self-absorption, non-resonant \ion{Fe}{II*} emission, and \ion{C}{II}] and [\ion{O}{II}] nebular emission lines.
The \FeII\ transitions occur in three multiplets\footnote{See \citet{2014ApJ...793...92T} or \citet{2015ApJ...815...48Z}  for energy level diagrams.}. In the \ion{Fe}{II} UV1, UV2, and UV3 multiplets, a photon can be re-emitted either through a resonant transition to the ground state, which produces emission infilling, or through a non-resonant transition to an excited state in the lower level, in which case the emission occurs at a slightly different wavelength.
We investigate the integrated absorption and emission profiles, focusing first on the resonant absorption and emission properties (Section~\ref{section:res:abs}), then on the non-resonant emission properties (Section~\ref{section:nonres:em}).

\subsection{Resonant Fe and Mg profiles}
\label{section:res:abs}

Figure~\ref{absFits} compares the velocity profiles of each of the individual \ion{Fe}{II}, \ion{Mg}{II}, and \ion{Mg}{I} transitions relative to the galaxy systemic redshift. The self-absorption profiles are asymmetric, with the strongest component centered on the galaxy systemic redshift, and a significant blue wing extending to $-800$~km~s$^{-1}$. 
We fit these profiles simultaneously with VPFIT\footnote{\url{http://www.ast.cam.ac.uk/~rfc/vpfit.html}} v10, using several components and requiring each to have the same redshift and Doppler parameter across the different transitions.  The absorptions are well-fitted with three components at redshifts $1.28514 \pm 0.00021$, $1.28752 \pm 0.00009$, and $1.29024 \pm 0.00006$, corresponding to shifts of  $-660 \pm 28$~km~s$^{-1}$, $-349 \pm 12$~km~s$^{-1}$ and  $+8.5\pm6.5$~km~s$^{-1}$ relative to the galaxy systemic velocity.  
Table~\ref{table:eqW} summarizes the total rest-frame equivalent widths for each transition, calculated both from the fit and directly from the flux.

Globally, the \FeII\ resonant transitions in Figure~\ref{absFits} reveal several key features: 
(1) the \FeII\ profiles are very similar to one another, and
(2) the strongest component is roughly centered at the galaxy systemic redshift. 
As \cite{2011ApJ...734...24P} first demonstrated, emission infilling in resonant absorption lines can alter doublet ratios and mimic partial coverage. 
However, here we find that emission infilling does not play a significant role in this galaxy for the following two qualitative arguments. 

First, while strong emission infilling would produce clear P-cygni profiles (which are not observed), moderate amounts of emission infilling would cause a blue-shift to the centroid of the absorption, an effect commonly seen in stacked spectra \citep[e.g.,][]{2015ApJ...815...48Z} or individual cases \citep{2011ApJ...728...55R,2013ApJ...770...41M}.
None of the absorptions in the galaxy ID\#13 spectrum (Figure~\ref{absFits}) have blue-shifted centroids. 

Second, because \FeII\ has multiple channels to re-emit the photons (through resonant and non-resonant transitions), the degree of infilling for a particular  \FeII\ absorption line depends on the likelihood of re-emission through the different channels within a multiplet. Purely resonant transitions, such as \ion{Mg}{II} and \FeII\,$\lambda2383$, are the most sensitive to emission infilling.
\citet{2015ApJ...815...48Z} demonstrated that the \FeII\ resonant absorptions that are the least (most) affected by emission infilling are \FeII\,$\lambda2374$ (\FeII\,$\lambda2600$ and \FeII\,$\lambda2383$) respectively. Figure~\ref{absFits} shows that the \FeII\,$\lambda2374$, $\lambda2600$ and $\lambda2383$ absorption profiles are all very similar for the galaxy ID\#13. The lack of blue-shifted centroids and the consistent absorption profiles argue strongly against the presence of detectable emission infilling in this galaxy.

We quantify (and put a limit on) the global amount of infilling using the method proposed by \citet{2015ApJ...815...48Z}, which consists of
comparing the observed rest-frame equivalent widths of the resonant lines to those seen in intervening quasar spectra (see their Figure 12). Using the averaged rest-frame equivalent widths of resonant \FeII\ and \ion{Mg}{II} absorptions from a stacked spectrum of $\sim 30$ strong \ion{Mg}{II} absorber galaxies at $0.5<z<1.5$ from \citet[][their Table 7]{DuttaR_17a}, we find that our data is consistent with no emission infilling.
Our data could allow for at most $<$0.8~\AA\ ($<$1.8~\AA) of infilling for \FeII\,$\lambda2600$ (\FeII\,$\lambda2383$), the two transitions most susceptible to infilling \citep{2015ApJ...815...48Z}.
This means that at most 22\%\ (55\%) of these absorptions could be affected by infilling and that the impact on the other \FeII\ absorptions is even smaller. 

Similarly, we separately estimate the amount of infilling for each of the three sub-components shown in Figure~\ref{absFits} (Table~\ref{table:eqW}). We are unable to put constraints on the weak component `A', but the blue-shifted component `B' at $-350$~km s$^{-1}$ does not allow for emission infilling that would increase the \FeII\,$\lambda2383$ equivalent width by more than 10\%. The component `C' at the galaxy systemic redshift allows for the largest amount of emission infilling with 60\%\ corrections for \FeII\,$\lambda2600$ and \FeII\,$\lambda2383$, 40\%\ for \FeII\,$\lambda2344$ and 20\%\ for \FeII\,$\lambda2586$. As we discuss later in Section~\ref{section:outflowrate}, the blue-shifted galactic wind component (`B') appears to be less affected by emission infilling than the systemic component associated with the galaxy ISM (`C').

We end this section by mentioning that, as we will argue in section~\ref{section:outflowrate}, the \FeII\ and \ion{Mg}{II} gas is likely optically thick. The absorptions ought to be saturated, and the reason we do not observe fully absorbed profiles is either due to a partial covering fraction (rather than emission infilling) or more likely to the low spectral resolution.
As we will show in the next section, the non-resonant \FeIIs\ emission pattern is also consistent with optically thick gas.

\begin{table*}[!p]
\centering
\caption{Emission and absorption rest-frame equivalent width and flux values}
\label{table:trans}
\begin{tabular}{lccccccc}  
\hline\hline        
Multiplet &   $\lambda$ & $\rm E_{high}$ & $\rm E_{low}$ & $J$ & $A_{ul}$ & $W_0$ & Flux \\
  & \AA & $\rm cm^{-1}$  & $\rm cm^{-1}$ &     & $\rm s^{-1}$ & \AA & $\rm 10^{-18}$~\flux \\
  (1) & (2) & (3) & (4) & (5) & (6) & (7) & (8) \\
\hline
\multirow{5}{*}{\ion{Fe}{II} UV1} &2600.17 & 38458.98 &   0.00 & 9/2$ \leftarrow $9/2 &  Absorption & $ 4.28 \pm  0.25$ & ... \\
                 & 2626.45 &38458.98 & 384.79 & 9/2$ \rightarrow $7/2 &  3.41E+07 & $ -0.93 \pm  0.13$ & $2.67 \pm  0.43$ \\
                 & 2586.65 &38660.04 &   0.00 & 7/2$ \leftarrow $9/2 &  Absorption & $ 3.14 \pm  0.26$ & ... \\
                 & 2612.65 &38660.04 & 384.79 & 7/2$ \rightarrow $7/2 &  1.23E+08 & $ -0.53 \pm  0.15$ & $1.47 \pm  0.49$ \\
                 & 2632.11 & 38660.04 & 667.68 & 7/2$ \rightarrow $5/2 & 6.21E+07 & $> -0.27$ & $< 0.78 \pm 0.42$~~~~  \\  
\hline
   & 2382.76 &41968.05 &   0.00 & 11/2$ \leftarrow $9/2~ &  Absorption\tablefootmark{a} & $ 3.27 \pm  0.22$ & ... \\
 {\ion{Fe}{II} UV2}               & 2374.46 & 42114.82 &   0.00 &  ~9/2$ \leftarrow $9/2 &  Absorption & $ 2.20 \pm  0.22$ & ... \\
                 & 2396.36 & 42114.82 & 384.79 &  ~9/2$ \rightarrow $7/2 &  2.67E+08 & $ -0.84 \pm  0.17$ & $2.37 \pm  0.49$ \\
\hline             
  &  2344.21 &42658.22 &   0.00 &  7/2$ \leftarrow $9/2 & Absorption & $ 3.47 \pm  0.24$ & $\cdots$ \\
 {\ion{Fe}{II} UV3}                & 2365.55 & 42658.22 & 384.79 &  7/2$ \rightarrow $7/2 &  5.90E+07 & $ -0.42 \pm  0.15$ & $1.22 \pm  0.48$ \\
                 & 2381.49 & 42658.22 & 667.68& 7/2$\rightarrow$5/2 & 3.10E+07\tablefootmark{b} & $\cdots$  & $\cdots$ \\
\hline 
\hline
\multirow{5}{*}{\ion{C}{II}]} & 2324.21 & 43025.3 & 0.00 & 3/2$ \rightarrow $1/2 &  ... & \multirow{5}{*}{$-1.03 \pm  0.18$} & \multirow{5}{*}{$2.83 \pm 0.50$}  \\   
                              & 2325.40 &43003.3 & 0.00 &  1/2$ \rightarrow $1/2 &  ... &  & \\   
                              & 2326.11 &43053.6 & 63.42 & 5/2$ \rightarrow $3/2 &  ... &  & \\   
                              & 2327.64 &43025.3 & 63.42 & 3/2$ \rightarrow $3/2 &  ... &  & \\   
                              & 2328.83 &43003.3 & 63.42 & 1/2$ \rightarrow $3/2 &  ... &  & \\

\hline
\multirow{2}{*}{[\ion{O}{II}]} & 3727.10 & 26830.57 & 0.00 & 3/2 $\rightarrow $3/2 &  ... & \multirow{2}{*}{$-48.98 \pm 0.29$} & \multirow{2}{*}{$133.46 \pm 0.80$}~~~~ \\   
              & 3729.86 & 26810.55 & 0.00 & 5/2 $\rightarrow $3/2 &  ... &  &  \\
\hline 
\end{tabular}
\tablefoot{Column (1): Transition name.
Column (2): Transition wavelength.
Column (3): Upper energy level.
Column (4): Lower energy level.
Column (5): Level total angular momentum quantum number $J$.
Column (6): Einstein $A_{ul}$ coefficient for spontaneous emission.
Column (7): Rest-frame equivalent width.
Column (8): Line flux.
}
\tablefoottext{a}{\FeII$\lambda$2382 is a pure resonant transition with no associated \FeIIs\ emission.}
\tablefoottext{b}{\FeIIs$\lambda$2381 emission is blended with \FeII$\lambda$2382 absorption.}
\end{table*}

\begin{table*}[!p]
\caption{Summary of two-dimensional morphological analysis}  
\label{table:galfit}
\centering                                      
\begin{tabular}{l c c c}          
\hline\hline   
           & \multirow{2}{*}{\ion{Fe}{II}*} & Stellar & \multirow{2}{*}{[\ion{O}{II}]} \\
           &                                & Continuum &  \\
\hline   
Axis ratio   & $0.57 \pm 0.01$ & $0.90 \pm 0.02$ & $0.85 \pm 0.01$ \\ 	
$R_{1/2}$ (arcsec) & \rhalfFeiiarcsec & \rhalfContarcsec & \rhalfOiiarcsec \\
$R_{1/2}$ (kpc)    & \rhalfFeiikpc & \rhalfContkpc & \rhalfOiikpc \\
\hline              
\end{tabular}
\end{table*}

\subsection{Non-resonant emission}
\label{section:nonres:em}

Figure~\ref{emissFits} shows the non-resonant transitions \ion{Fe}{II}*\,$\lambda2365$, $\lambda2396$, $\lambda2612$, and $\lambda2626$ that we detect in the MUSE HDFS galaxy ID\#13 1D spectrum at $2.5\sigma - 6\sigma$ significance. 
No \ion{Fe}{II}*\,$\lambda2632$ emission is detected  (Figure~\ref{fig:spec1D}).
The fluxes in the non-resonant transitions \ion{Fe}{II}*\,$\lambda2365$, $\lambda2396$, $\lambda2612$, $\lambda2626$ transitions
are $1.2 - 2.4 - 1.5 - 2.7 \times10^{-18}$~\flux, respectively.
Table~\ref{table:trans} gives the emission peak fluxes and rest-frame equivalent widths measured for all of the \ion{Fe}{II}* transitions. 
These flux ratios of $0.5:1.0:0.6:1.0$ are consistent with the expectation ($0.66:1.0:0.66:1.0$) for optically thick gas discussed in \citet{2014ApJ...793...92T}. In the optically thin regime, the flux ratios should be on the order of $\sim 1$.

Regarding the non-detection of \FeIIs\,$\lambda2632$, we note that this transition is usually not detected in stacked spectra
\citep{2012A&A...539A..61T, 2013ApJ...774...50K, 2014ApJ...793...92T, 2015ApJ...815...48Z}, except for in the \citet{2012ApJ...759...26E} stacked spectrum, but that it is observed in the other individual cases \citep{2011ApJ...728...55R,2013ApJ...770...41M}.
\cite{2014ApJ...793...92T} explore whether underlying stellar absorption suppresses the \ion{Fe}{II}*\,$\lambda2632$ emission in their stacked spectra.
However, for this starburst galaxy, the F- and G-type stars that produce the underlying absorption are unlikely to significantly contribute to the stellar continuum.

We perform a joint Gaussian fit to the four non-resonant \FeIIs\ emission peaks and find that they appear symmetric and centered on the galaxy systemic redshift measured from [\ion{O}{II}]\,$\lambda\lambda3727,3729$ (Figure~\ref{emissFits}). 
This is in contrast to \citet{2015ApJ...815...48Z},
who found that the \FeIIs\ emission from their stacked spectrum of 8,600 galaxies is slightly asymmetric, and in contrast to \cite{2011ApJ...728...55R}, who observed \ion{Fe}{II}* emission peaks that are slightly ($\sim 30$~km~s$^{-1}$) redshifted relative to the nebular emission lines.

\section{Morphology of the \ion{Fe}{II}* Emission}
\label{section:2D}

\begin{figure*}[!p]
\resizebox{\hsize}{!}{\includegraphics{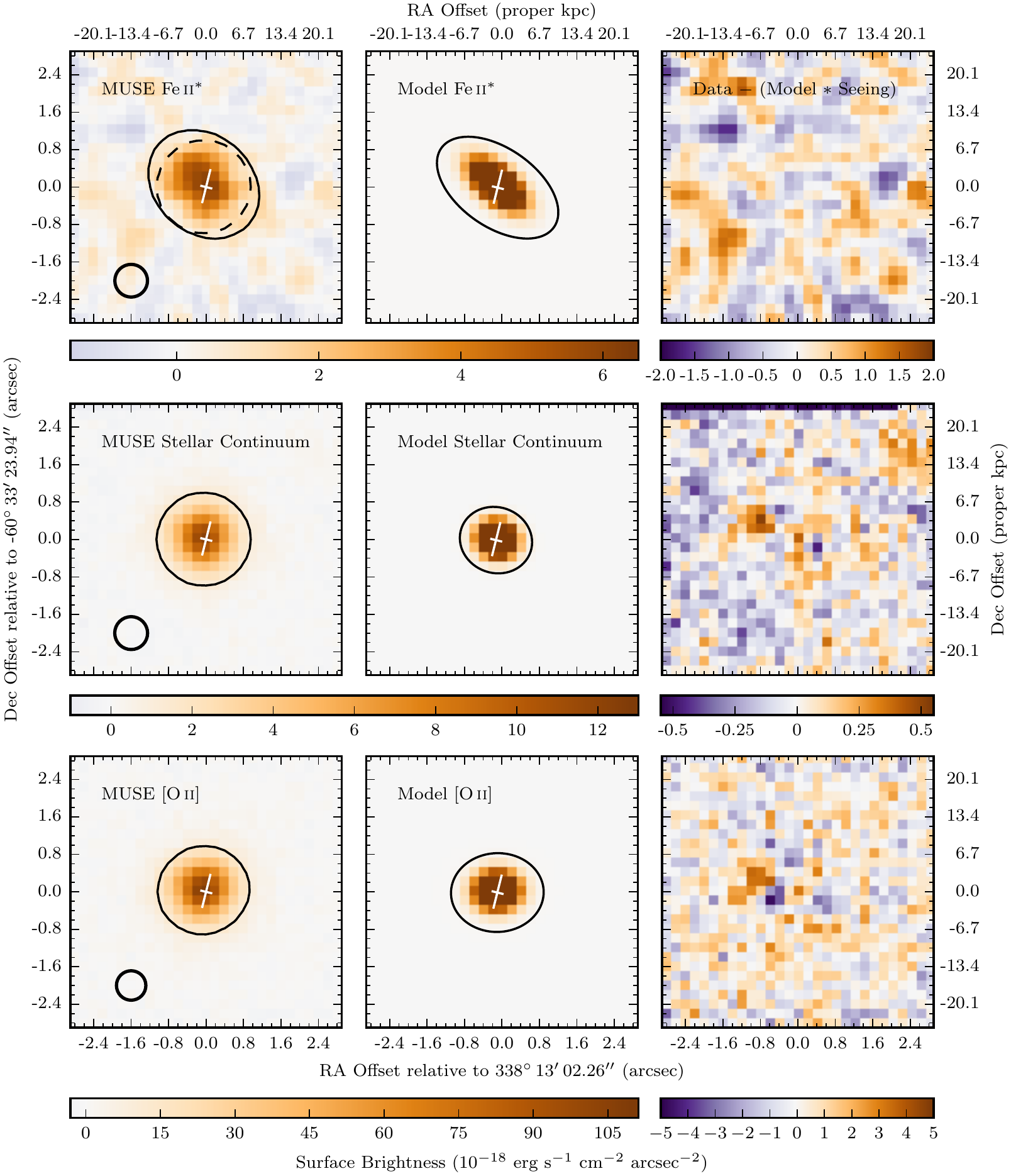}}
\caption{{\it Left panels:} Surface brightness maps of \ion{Fe}{II}* (top), stellar continuum (middle), and [\ion{O}{II}] emission (bottom) from pseudo narrow band images  (see text). Solid contours represent $1/10$ of the maximum surface brightness.
The dashed line in the top left panel correspond to the stellar continuum contour from middle left panel.
The small black circles  represent the seeing at the emission wavelength.
{\it Middle panels:} Surface brightness maps of the intrinsic emission from an exponential disk model `deconvolved' from the seeing. Ellipses in the middle column are drawn using the model parameters and have a size that corresponds to the half-light radii (see Table~\ref{table:galfit}).
{\it Right panels:} Maps of  the residuals between the observed data and the intrinsic model convolved with the seeing.    The white crosses indicate the galaxy major and minor axes from the \cite{2016A&A...591A..49C} kinematic analysis. The \ion{Fe}{II}* emission map is more extended than both the stellar continuum or the \OII\ emission.}
\label{sbMaps}
\end{figure*}

\begin{figure*}[!ht]
\resizebox{\hsize}{!}{\includegraphics{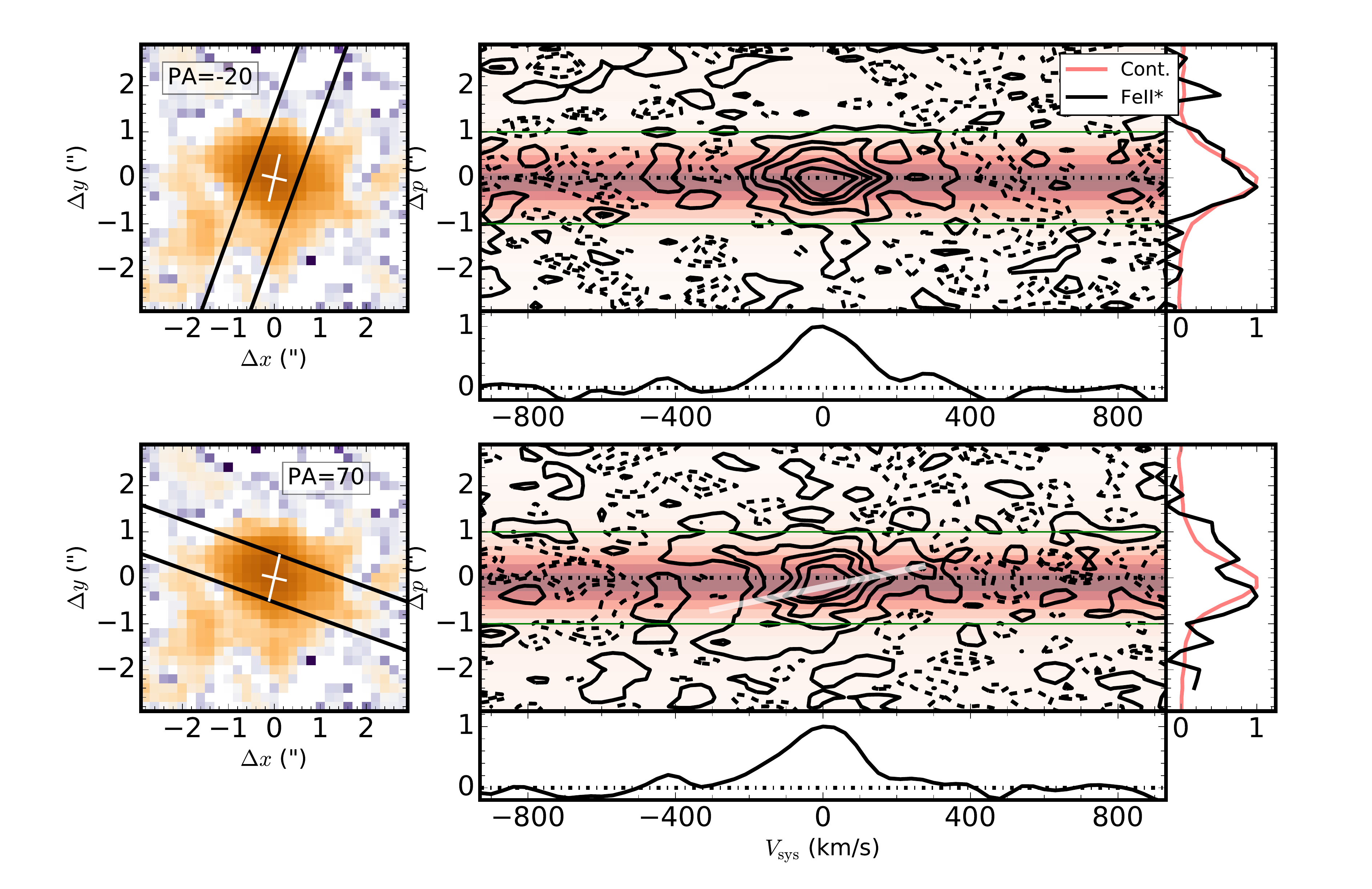}}
\caption{{\it Left panels:} Narrow band images of the \ion{Fe}{II}* emission optimally extracted from the MUSE cube shown with a 1\arcsec-wide slit oriented -20$^{\circ}$ (top) and +70$^{\circ}$ (bottom). The +70$^{\circ}$ slit orientation in the bottom panels follows the \ion{Fe}{II}* major kinematic axis. The white crosses indicate the galaxy major and minor axes from the \cite{2016A&A...591A..49C} kinematic analysis of [\ion{O}{II}]. 
{\it Right panels:} Position-velocity diagrams of the \ion{Fe}{II}* emission from a slit oriented -20$^{\circ}$ (top) and +70$^{\circ}$ (bottom). Zero velocity is relative to the galaxy systemic redshift, $z = $\redshift, measured from the [\ion{O}{II}] emission. Solid black contours trace \ion{Fe}{II}* flux levels at 2, 6, 10, 14, and $18\times 10^{-20}$~erg~s$^{-1}$~cm$^{-2}$.  Dashed black contours show the negative flux levels at -2 and -4~$\times 10^{-20}$~erg~s$^{-1}$~cm$^{-2}$. The red gradient indicates the continuum intensity.
The right side panel of each p-v diagram compares the spatial profile of the \ion{Fe}{II}* emission (black) with that of the stellar continuum (red). 
The bottom sub-panel of each p-v diagram shows the total 1D flux spectrum integrated across the spatial region between the solid green horizontal lines in the p-v diagram. 
In the bottom p-v diagram, the white solid line follows the velocity gradients. This panel also reveals a `C'-shape pattern in the blue wing of the \FeIIs\ emission extending to -400~km~s$^{-1}$.
 The blue wing of the \ion{Fe}{II}* emission is more pronounced along the slit orientation corresponding to the minor-axis (PA$= +70^{\circ}$). 
}
\label{pvPlots}
\end{figure*}

In this section, we investigate whether the \ion{Fe}{II}* emission has a similar spatial extent and morphology as the stellar continuum and the [\ion{O}{II}]\,$\lambda\lambda3727,3729$ emission.

For the \ion{Fe}{II}* emission, first we produce a sub-cube of size $1.5\arcsec \times 1.5\arcsec$ for each of the four emission lines and transform the wavelength axis to velocity space. We interpolate each sub-cube to the same velocity scale with pixels of 30~km~s$^{-1}$ that span $\pm930$~km~s$^{-1}$ and zero velocity at the galaxy systemic redshift, $z = 1.2902$. 

We subtract the continuum and combine the four sub-cubes.
To estimate the stellar continuum, we use the mean value from two regions redwards of the \ion{Fe}{II}* emission peaks at $\sim \lambda2425$ and $\sim \lambda2700$~\AA\ that span 115~\AA\ and 300~\AA\ respectively.
The continuum pseudo narrow
band image shown in Figure~\ref{sbMaps} (middle left) is from the mean of these two continuum regions, which have a flat slope.
 
From the combined \ion{Fe}{II}* emission velocity cube, we then extract a narrow band (NB) image by summing 13 pixels ($\pm 390$~km~s$^{-1}$). 
The top left panel of Figure~\ref{sbMaps} shows the pseudo-narrow band \ion{Fe}{II}* image with $2 \times 2$ smoothing, and we use this image for the analysis. 
For comparison, we also tested an automated extraction with the CubExtractor software (Cantalupo et al.\, in prep.), which selects connected volume pixels (voxels) that are above a specified SNR threshold (2.7 was optimal in our case) to produce optimally extracted images, as in \cite{2016ApJ...831...39B}.
Our morphological results are independent of the method used to produce the \ion{Fe}{II}* NB image.

Similarly, we create the [\ion{O}{II}] pseudo-narrow band image from a $30 \times 30$ pixel ($1.5\arcsec \times 1.5\arcsec$) sub-cube that spans 18 spectral pixels (22.5~\AA) to cover
the $\lambda\lambda 3727,3729$ doublet. Again, we subtract the continuum estimated between $\sim3550 - 3600$~\AA\ to obtain the [\ion{O}{II}] surface brightness map shown in the bottom left panel of Figure~\ref{sbMaps}.

The \ion{Fe}{II}* map in Figure~\ref{sbMaps} is the first two-dimensional spatial map of the \FeIIs\ non-resonant emission in a individual galaxy at intermediate redshift. 
Previous studies have looked for signatures of extended \ion{Fe}{II}* emission in stacked spectra \citep{2012ApJ...759...26E,2014ApJ...793...92T}.
In a stacked spectrum from 95 star-forming galaxies at $1 < z < 2$, \cite{2012ApJ...759...26E} found that the \ion{Fe}{II}*\,$\lambda2626$ emission line is slightly more spatially extended that the stellar continuum.
\cite{2014ApJ...793...92T} performed a similar analysis with 97 star-forming galaxies at $1 \lesssim z \lesssim 2.6$, but were not able to spatially resolve the \ion{Fe}{II}* emission.

Thanks to the sensitivity of MUSE, we are able to address whether \ion{Fe}{II}* is more extended than the continuum and to characterize the \ion{Fe}{II}* emission morphology for the first time.
The top left panel of Figure~\ref{sbMaps}) shows that the extended \ion{Fe}{II}* emission (solid contour) appears to be more extended than the continuum (dashed contour) and has a privileged direction. Comparing the \ion{Fe}{II}* emission position angle with the kinematic axis of the galaxy, indicated with crosses, shows that the \ion{Fe}{II}* is more extended along the minor kinematic axis of the galaxy.

To quantify the extent of the \ion{Fe}{II}*, stellar continuum, and [\ion{O}{II}]\,$\lambda\lambda3727,3729$ emission, we use a custom Python MCMC algorithm to fit each of the surface brightness maps in the left column of Figure~\ref{sbMaps} with a Sersic profile. The fit provides us with intrinsic parameters and with an intrinsic model of the emitting region, i.e. \textit{deconvolved from the seeing}, because we convolve the Sersic profile with the actual PSF taken from the brightest star in the same data cube, MUSE HDFS ID\#1 \citep[see][]{2015A&A...575A..75B}, across wavelengths corresponding to the galaxy emission lines~\footnote{The PSF can be approximately described by a Moffat profile with FWHM 0.70\arcsec\ (0.63\arcsec) at the \ion{Fe}{II}* and stellar continuum emission ([\ion{O}{II}] emission) wavelengths, which corresponds to a half-light radius of 0.50\arcsec\ (0.44\arcsec).}. 
In practice, we fix the Sersic index $n$  to $n=1$ or $n=0.5$ because the Sersic index $n$ is unconstrained~\footnote{The Sersic $n$ index is unconstrained because the seeing radius is much larger than the emission. Indeed, the seeing radius  is FWHM/2=0.35\arcsec,   corresponding to  $R_{1/2}\approx$0.5\arcsec\ for a Moffat profile, whereas the galaxy's intrinsic half-light radius $R_{1/2}$ is only $\approx0.3$\arcsec.}. The size estimate, $R_{1/2}$, is nonetheless robust and independent of the Sersic index $n$, since it is determined empirically from the flux growth curve, an integrated quantity.

Table~\ref{table:galfit} summarizes the results from this analysis and Figure~\ref{sbMaps} (middle column) shows the modeled profiles for $n=1$ for the \ion{Fe}{II}*, stellar continuum, and [\ion{O}{II}] emission. The right column of Figure~\ref{sbMaps} gives the residual maps, which are the difference between the observed data and the intrinsic model convolved with the seeing.

The stellar continuum emission (Figure~\ref{sbMaps}, middle row) appears round and compact. The intrinsic emission from the exponential disk fit yields an inclination of $28 \pm 3^{\circ}$ and a half-light radius, $R_{1/2}$, of around \rhalfContarcsec\arcsec\ (\rhalfContkpc~kpc). 
These continuum emission properties from MUSE are comparable to the measurements from HST images discussed in Section~\ref{sect:galProp} and shown in Table~\ref{table:galProps}.
The [\ion{O}{II}]\,$\lambda\lambda$3727,3729 emitting region has the same morphology but is slightly more extended than the stellar continuum with $R_{1/2,\OII} =$ \rhalfOiiarcsec\arcsec\ (\rhalfOiikpc~kpc). The corresponding star formation rate surface density is \sigSFR.

The \ion{Fe}{II}* emission has a morphology and physical extent that are different from the stellar continuum and [\ion{O}{II}] emission. The intrinsic \ion{Fe}{II}* emission is more elliptical with an axis ratio of $b/a=0.57$, compared to the rounder continuum and [\ion{O}{ii}] emission, which both have $b/a\simeq0.9$.
The \ion{Fe}{II}* emission is elongated along the direction (${\rm PA} \approx +60^\circ$) that roughly corresponds to the galaxy minor kinematic axis (${\rm PA} \approx +75^\circ$, Table~\ref{table:galProps}). 
Moreover, the intrinsic half-light radius of the \ion{Fe}{II}* emission is $R_{1/2,\,\FeIIs}=$~\rhalfFeiiarcsec \arcsec, i.e. about 50\%\ larger than that of the stellar continuum. In other words, the \ion{Fe}{II}* half-light radius, $R_{1/2,\,\FeIIs}=$~\rhalfFeiikpc~kpc, extends $\gtrsim 1.5$~kpc beyond the stellar continuum and the [\ion{O}{II}] emission, which both have $R_{1/2} \approx 2.5$~kpc (Table~\ref{table:galfit}).
This is apparent from comparing the extent of the \ion{Fe}{II}* emission (solid contour) to the continuum emission (dashed contour) in the top left panel of Figure~\ref{sbMaps}.
See Table~\ref{table:galfit} for the emission properties.

\section{Kinematics of the \ion{Fe}{II}* emission}
\label{section:kinematics}

In this section, we investigate whether it is possible to trace the kinematics of the \ion{Fe}{II}* emission.
To do so, we visually inspected the velocity cube produced in the previous section and found that the kinematic major axis from the \ion{Fe}{II}* emission follows a PA of about $70$~deg, which happens to correspond roughly to the galaxy minor kinematic axis.
Figure~\ref{pvPlots} shows $p-v$ diagrams for this $70$~deg slit orientation (bottom row) and for a slit oriented at $-20$~deg (top row). In both cases, the slit width is 1\arcsec. Following the peak of the \FeIIs\ emission, we see that the \ion{Fe}{II}* emission has a velocity gradient along the galaxy minor kinematic axis. 
The white solid line in the bottom $p-v$ diagram guides the eye along this velocity gradient.
Black contours trace the \ion{Fe}{II}* emission, and the red shaded area indicates the continuum.

Figure~\ref{pvPlots} reveals two additional results.
First, the \ion{Fe}{II}* emission shows an extended blue-wing, which is more pronounced in the $PA = +70$ profile. Secondly, the blue-side of the \ion{Fe}{II}* emission contours in the bottom panel, with the slit oriented at $+70$~deg, shows a C-shaped pattern. The contours extend to $-400$~km~s$^{-1}$ near $+1$\arcsec\ and $-1.5$\arcsec, but decrease to $-200$~km~s$^{-1}$ in between. This C-shape pattern is characteristic of a hollow conical emission, as could be expected from an outflow.

\section{Discussion}
\label{sect:discuss}

\begin{figure*}[!ht]
\resizebox{\hsize}{!}{\includegraphics{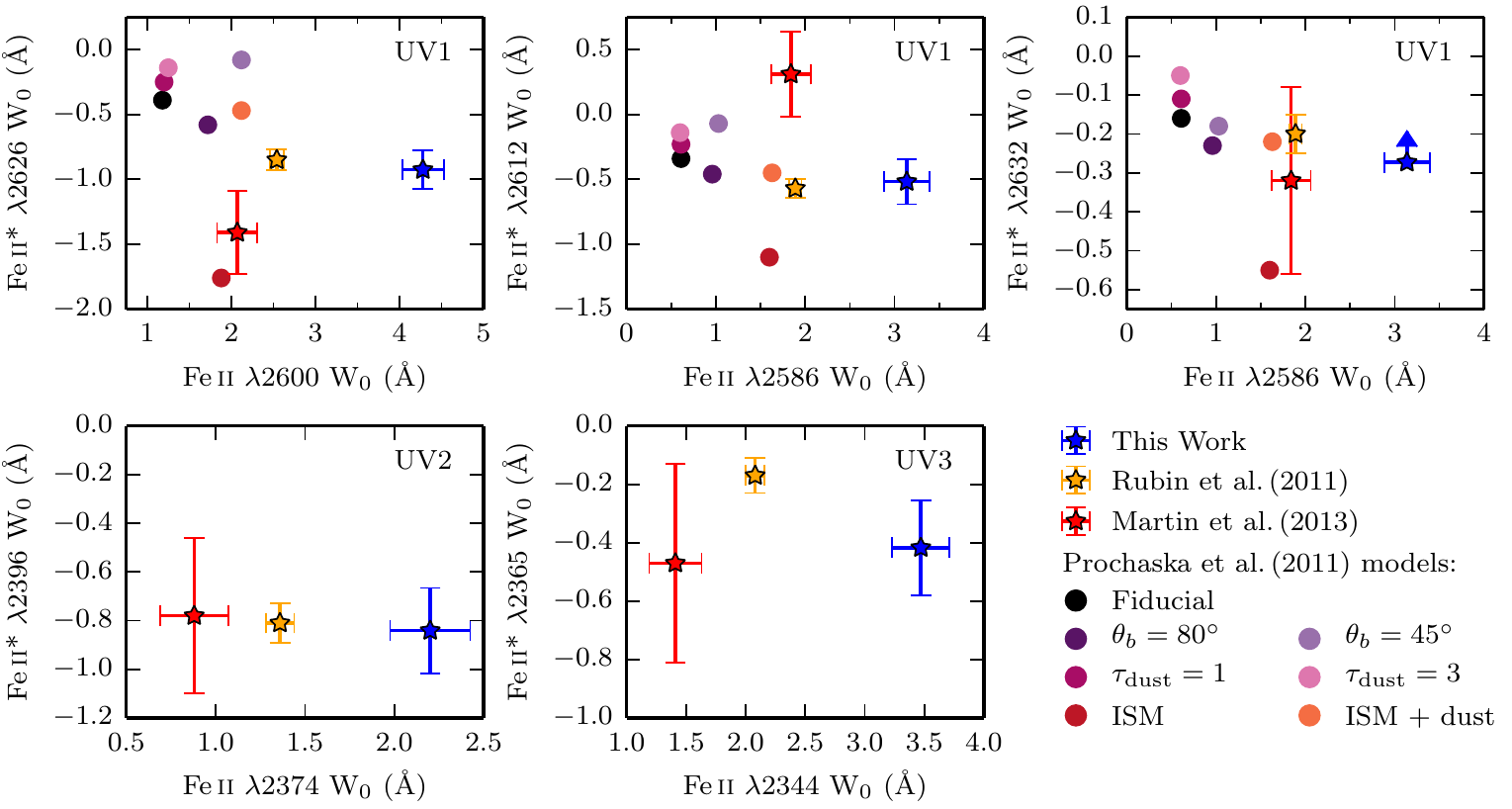}}
\caption{Observed and modeled rest-frame equivalent widths from the \ion{Fe}{II} UV1, UV2, and UV3  multiplet transitions. Star-shaped points indicate values measured from the three direct detections: this work (blue), \cite{2011ApJ...728...55R} (yellow), and \cite{2013ApJ...770...41M} (red).
Circular points in the top panels come from radiative transfer model predictions for the UV1 multiplet \citep{2011ApJ...734...24P}. Colors represent variations to the fiducial isotropic outflow model (black) that test different opening angles ($\theta_b$) for bi-conical outflows, dust opacities, and include an ISM component. See \cite{2011ApJ...734...24P} for additional model variations. Model predictions for the \ion{Fe}{II} UV2 and UV3 multiplets are not currently available.}
\label{eqWcomp}
\end{figure*}

From deep MUSE observations of the HDFS, we identify a spatially-resolved galaxy (ID\#13) at $z=1.2902$ that has a low inclination ($i = 33^{\circ}$), an orientation that may favor detecting galactic outflows in emission \citep{2014ApJ...794..156R}.
This galaxy has a star formation rate of \sfr, which places it in the starburst category \citep[See][their Figure 3]{2016A&A...591A..49C}.  
Its star formation rate surface density is \sigSFR, well above the threshold for galactic winds \citep{2012ApJ...761...43N, 2014ApJ...784..108B}.
The star formation rate and stellar mass are nearly identical to those of two galaxies at $z = 0.694$ \citep{2011ApJ...728...55R} and $z = 0.9392$ \citep{2012ApJ...760..127M} that both show direct evidence of galactic winds from blueshifted absorptions, redshifted \ion{Mg}{II} emission, and non-resonant \ion{Fe}{II}* emission.

In the integrated 1D MUSE spectrum, we detect non-resonant \ion{Fe}{II}* emission but no apparent resonant \ion{Fe}{II} and \ion{Mg}{II} emission, differentiating this galaxy somewhat from the two previous examples. 
We also obtain the first direct detection of spatially extended \ion{Fe}{II}* emission (Section \ref{section:2D}) from an individual galaxy by stacking the \ion{Fe}{II}* transitions (Figure~\ref{sbMaps}).
We discuss here (in Section~\ref{section:implications}) the implications for these results in the context of radiative transfer wind models, since the emitting gas is likely entrained in a galactic-scale outflow.

The strong, asymmetric \ion{Fe}{II} and \ion{Mg}{II} absorptions in the 1D galaxy spectrum, which have blueshifted components at $-660$~km~s$^{-1}$ and $-350$~km~s$^{-1}$ relative to the systemic redshift (Figure \ref{absFits}), are a clear signature of outflowing gas. 
In Section~\ref{section:outflowrate}, we will estimate the mass outflow rate and compare it to the galaxy SFR.

\subsection{Implications for outflow models}
\label{section:implications}

The MUSE surface brightness maps (Figure~\ref{sbMaps}) reveal that the \ion{Fe}{II}* emission has a more elliptical shape than the stellar continuum and the [\ion{O}{II}] emission. 
Detecting \ion{Fe}{II}* emission that is more extended along one axis suggests that the outflow is not isotropic.
Isotropic outflows are, however, the fiducial geometry for radiative transfer and semi-analytic wind models \citep{2011ApJ...734...24P, 2015ApJ...801...43S}.
Comparing these models and their variations with direct detections, such as the MUSE HDFS galaxy ID\#13, can help to interpret the observations while motivating additional refinements to the models.

The radiative transfer and semi-analytic models of galactic outflows from \citet{2011ApJ...734...24P} and \citet{2015ApJ...801...43S} both predict concurrent resonant and non-resonant emission, because absorbed photons
can be re-emitted via either non-resonant or resonant transitions for multiplets \citep{2014ApJ...793...92T,2015ApJ...815...48Z}.
The spectrum of the HDFS galaxy ID\#13 (Figures~\ref{fig:spec1D}~and~\ref{absFits}) shows strong \FeII\ and \ion{Mg}{II} absorptions (with total rest-frame equivalent widths from 3 to 5 \AA), but no evidence for P-cygni profiles and globally small amounts of possible emission infilling as discussed in Section~\ref{section:res:abs}. We next discuss whether this apparent lack of \FeII\ resonant emission occurs in other galaxies, and how to potentially reconcile the models with such data.

Like the HDFS ID\#13 galaxy, the two previously published direct detections of \ion{Fe}{II}* emission  \citep{2011ApJ...728...55R,2013ApJ...770...41M} do not have \FeII\ P-cygni profiles, although they might have moderate resonant \FeII\ emission infilling. 
Unlike the HDFS ID\#13 galaxy, these galaxies have strong  P-cygni \ion{Mg}{II} profiles.
The published composite spectra \citep{2012A&A...539A..61T, 2012ApJ...759...26E, 2013ApJ...774...50K, 2014ApJ...793...92T, 2015ApJ...815...48Z} show non-resonant \ion{Fe}{II}* emission without obvious resonant \ion{Fe}{II} emission as P-cygni profiles.
Among these composite spectra, only \citet{2012ApJ...759...26E} and \citet{2013ApJ...774...50K} reveal \ion{Mg}{II} P-cygni profiles.

In order to investigate the origin of the apparent lack of resonant \FeII\ emission,
in Figure~\ref{eqWcomp} we compare rest-frame equivalent width measurements from the MUSE HDFS ID\#13 galaxy with predictions from the \citet{2011ApJ...734...24P} radiative transfer models, following \citet{2012ApJ...759...26E}.  In Figure~\ref{eqWcomp}, we also include the two previously published direct detections of \FeIIs\ emission from \citet{2011ApJ...728...55R} and \citet{2013ApJ...770...41M}.
Each panel pairs a non-resonant \ion{Fe}{II}* emission transition with its corresponding resonant \ion{Fe}{II} absorption transition from within a multiplet.

\cite{2011ApJ...734...24P} produce models for the \ion{Fe}{II} UV1 multiplet and \ion{Mg}{II}\,$\lambda\lambda2797,2803$ doublet to explore how varying model geometries and physical assumptions about the dust content, ISM contribution, gas density, and wind speeds impact the line profiles from the resonant and non-resonant transitions.
In nearly all of the tested models, the resonant transitions produce P-cygni profiles with blueshifted absorption and redshifted emission.
Varying each of the physical properties individually from the fiducial model is not sufficient to suppress the resonant \ion{Fe}{II} and \ion{Mg}{II} P-cygni profiles. Indeed, the only model that substantially suppresses the resonant emission combines an ISM component with dust extinction.  
In order to reproduce the observed profiles of the MUSE HDFS galaxy ID\#13, the models will need to simultaneously incorporate more properties.
To gain physical intuition for the impact of the individual properties, we now discuss varying the outflow geometry, dust extinction and ISM component affects the profiles.

In contrast to their fiducial model (black point in Figure~\ref{eqWcomp}), which assumes angular isotropy, \cite{2011ApJ...734...24P} also modeled bi-conical outflows (purple points), where the wind fills an opening angle into and out of the plane of the sky along the line of the sight to the galaxy. 
Collimating the outflow suppresses both the resonant and non-resonant emission, and for highly collimated outflows, absorption dominates the profile. 
The lack of resonant emission in the HDFS galaxy ID\#13 suggests that the outflow could be bi-conical and collimated.
However, since this geometry also suppresses the non-resonant \ion{Fe}{II}* emission, highly collimated wind models create a double-peaked \ion{Fe}{II}* profile that is not observed.

The \cite{2011ApJ...734...24P} fiducial model assumes no dust extinction, but they show that dust absorption (pink points in Figure~\ref{eqWcomp}) can have a strong impact on the line profiles. 
Increasing the amount of dust extinction suppresses the resonant emission slightly more than the non-resonant emission.
Adding $\tau_{\rm dust}=3$ to the fiducial model leaves a weak \FeII\ P-cygni profile, whereas the model with $\tau_{\text{dust}} = 10$ suppresses the resonant \FeII\ emission while leaving weak non-resonant \ion{Fe}{II}* emission. However, such copious dust extinction would extinguish the source by 15 magnitudes and make it unobservable. 
The moderate visual extinction for the galaxy ID\#13, A$_\text{V} = 1.20^{+0.59}_{-0.26}$~mag, suggests that dust extinction alone is not sufficient to explain the diminished resonant emission.

Adapting the outflow model to include gas that represents the ISM of the galaxy (red circular point in Figure~\ref{eqWcomp}), i.e., gas that is centralized and lacks a significant radial velocity, also produces line profiles with similarities to the galaxy ID\#13. 
Adding the ISM component increases the absorption around zero systemic velocity and suppresses the resonant \FeII\ emission.
Moreover, the ISM component can boost the \FeIIs\ emission.
For \ion{Fe}{II}, more resonant absorption due to the ISM component allows more photons to escape through non-resonant re-emission. Indeed, the non-resonant \ion{Fe}{II}* emission becomes $\sim 10$ times stronger than in the fiducial model. 
With suppressed \FeII emission but increased \FeIIs emission, an ISM component is essential to re-creating the observations from the galaxy ID\#13.

Finally, \cite{2011ApJ...734...24P} include dust extinction in the ISM component (orange point in Figure~\ref{eqWcomp}). Compared to the dusty wind model discussed earlier, this model suffers much more from dust extinction because the simple kinematic structure of the ISM allows multiple scattering events. For \ion{Mg}{II}, the photons are resonantly trapped in the dusty ISM. 
For \FeII, all of the emission lines diminish compared to the same model without dust, but the ratio between the \ion{Fe}{II}* emission to \FeII\ emission remains stronger than in the fiducial model. This model best describes the ID\#13 galaxy. Further exploring the same physical conditions with a bi-conical outflow may highlight subtleties that could indicate preferring one geometry over another.

To summarize, we suggest that a model combining a dusty ISM with a bi-conical outflow that has a moderate amount of dust opacity in the wind would be able to match the data for the HDFS galaxy ID\#13.

\subsection{Mass Outflow Rate Estimation}
\label{section:outflowrate}

To estimate the mass entrained in the outflow, we consider only the absorption components that 
are not affected by the ISM. Consequently, we exclude the component `C' at the systemic velocity (Figure~\ref{absFits}, Table~\ref{table:eqW}). 
The other two components are blueshifted by $-660$~km~s$^{-1}$ (`A') and $-350$~km~s$^{-1}$ (`B'), respectively. The wind component `B' at $-350$~km~s$^{-1}$ dominates the bulk of the mass flux given the equivalent width ratios between components `A' and `B.' As discussed in Section~\ref{section:res:abs}, the wind component `B' is the least affected by emission infilling (at or below the 10\%\ level), whereas the ISM component `C' is the most affected by emission infilling. Hence, emission infilling does not affect our estimate of the mass outflow rate from the wind component `B.'

Similar to \cite{2014ApJ...794..156R}, we estimate the mass outflow rate from:
\begin{equation}
\label{dMdt}
\dfrac{dM}{dt} \approx 1~{\rm M}_{\odot}~{\rm yr}^{-1} C_{f} \dfrac{N_{\rm flow}({\rm H})}{\rm 10^{20}~cm^{-2}} \dfrac{A_{\rm flow}}{45~\rm kpc^2} \dfrac{v}{\rm 300~km~s^{-1}} \dfrac{\rm 5~kpc}{D}
\end{equation}
where $C_f$ is the covering fraction of the outflowing gas, $N_{\rm flow}({\rm H})$ is the column density of hydrogen associated with the outflow, $A_{\rm flow}$ is the projected surface area of the outflow, $v$ is the outflow velocity, and $D$ is the physical distance the outflow extends from the galaxy center.

We estimate $N_{\rm flow}({\rm H})$ from the metal column densities $N$(Fe) and $N$(Mg).
Because VPFIT column densities and Doppler $b$ parameters are degenerate for optically thick lines,
we determine the metal column densities $N$(Fe) and $N$(Mg) from the equivalent width, $W_0$,
following
 \cite{1968dms..book.....S}, with an additional term for the covering fraction:
\begin{equation}
\label{Nmetal}
{\rm log}~N = {\rm log}\dfrac{W_0}{\lambda} - {\rm log}\dfrac{2F(\tau_0)}{\pi^{1/2} \tau_0} - {\rm log}\lambda f - {\rm log} C_f + 20.053 
\end{equation}
where $\tau_0$ is the optical depth at line center, $\lambda$ is the transition wavelength in \AA , and $f$ is the oscillator strength.
The optical depth $\tau_0$ is determined from the ratio of equivalent widths from two lines within the same multiplet, as in \cite{2009ApJ...692..187W, 2010ApJ...719.1503R}, which is referred to as the `doublet ratio' method.

For \ion{Mg}{II}, the oscillator strengths indicate that the equivalent width ratio is $2:1$ in the optically thin case. The \ion{Mg}{II} equivalent width ratio follows $F(2\tau_0)/F(\tau_0)$ for the transmission integral:
\begin{equation}
F(\tau_0) = \int_0^{+\infty} (1 - e^{-\tau_0~exp(-x^2)})dx
\end{equation}
From our measured \ion{Mg}{II} equivalent width ratio, $1.06 \pm 0.13$, we numerically solve for $\tau_{0,\,2803} \approx 240$. Both the equivalent width ratio and the high optical depth value indicate that \ion{Mg}{II} is saturated.

For \ion{Fe}{II}, we can calculate the optical depth for two different sets of transitions: $\tau_{0,\,2586}$ from $W_{0,\,2600}/W_{0,\,2586}$ and $\tau_{0,\,2374}$ from $W_{0,\,2382}/W_{0,\,2374}$. The optical depth ratios are $3.46:1$ and $10.22:1$ respectively, using the oscillator strength values from \cite{2003ApJS..149..205M}. After again solving numerically, the optical depth values are $\tau_{0,\,2586} = 3.53$ and $\tau_{0,\,2374} = 1.55$. We can therefore use these optical depth and equivalent width values to obtain a good estimate of the \ion{Fe}{II} column density, since \cite{1986ApJ...304..739J} find accurate column densities even for blended components that result from multiple clouds, as long as the optical depth in the weaker transition is $\tau_0 < 5$.

With knowledge of the optical depth, we can determine the covering fraction from the residual intensities between the zero level and the doublet lines \citep{2005ApJS..160...87R, 2009ApJ...696..214S, 2009ApJ...703.1394M, 2012ApJ...760..127M, 2014ApJ...794..156R}.
Using Equation~5 from \cite{2005ApJS..160...87R} with \ion{Mg}{II}, we find a covering fraction of at least 0.4. However, this formula ignores the instrument resolution, which could lead to a much higher covering fraction. To estimate a lower limit on the column density, we take $C_f=1$, as in \cite{2010ApJ...719.1503R}.

Applying Equation~\ref{Nmetal}, the column density measurements are $N(\ion{Mg}{II}\,\lambda2803) = 15.89$, $N(\ion{Fe}{II}\,\lambda2586) = 14.74$, and $N(\ion{Fe}{II}\,\lambda2374) = 14.76$. These measurements are in good agreement with the values from vpFit, $N$(\ion{Mg}{II})~$= 15.87 \pm 0.68$ and $N$(\ion{Fe}{II})~$ = 14.75 \pm 0.16$. 

From the metal column densities, in order to estimate the gas flow column density $N_{\rm flow}$(H), we use solar abundances \citep[with log(Mg/H)~$= -4.40$ and log(Fe/H)~$= -4.50$;][]{2009ARA&A..47..481A} and a dust depletion correction but no ionization correction, as in \cite{2011ApJ...728...55R}. To estimate the dust depletion factor, we use the \cite{2009ApJ...700.1299J} method to simultaneously fit for the depletion level using  the column densities of these two elements (Mg, Fe).  
The fit yields a global depletion factor of $F_\star = 1.25 \pm 0.39$, corresponding to $\delta$Fe of -2.60 dex and $\delta$Mg of -1.50 dex.
With these depletion corrections, the total gas column density is thus at least log~$N$(H)~$ \geq 21.76 \pm 0.48 - \log Z/Z_{\odot}$, given that we used solar abundances\footnote{Ionization corrections would further increase the column density, but they are small at this level.}.

We can estimate the projected area of the ouflow $A_{\rm flow}$ from the size of the
stellar continuum, since we detect \ion{Mg}{II} and \ion{Fe}{II} in absorption against the  continuum.  
The MUSE stellar continuum (Section~\ref{section:2D}) has an intrinsic half-light radius of   \rhalfContkpc~kpc. Because the spectrum is optimally extracted with a white-light image weighting scheme, the {\it effective} half-light radius of the extracted 1D spectrum is $R_{1/2,\rm eff}=\sqrt{2} \times R_{1/2,\star}$ or 3.3~kpc.  The stellar continuum therefore covers a surface area of $A_{\rm flow}~=~\pi R_{1/2,\star}^2 b/a~=~$30~kpc$^2$.

Finally, we must assume an effective or characteristic distance for the gas at -350 km/s
 with a total column of $\log N_{\rm flow}(H)>21.80$. For a mass-conserving flow, the gas closer to the galaxy will dominate the column density. However, outflowing
 gas moving at $-350$~km~s$^{-1}$ needs a few kpc ($1-5$) to accelerate to that speed \citep{MurrayN_11a}. 
Hence, we conservatively use an upper limit of $D<5$~kpc, as in \cite{2014ApJ...794..156R}, which leads to an outflow rate of  $>45$~M$_{\odot}$~yr$^{-1}$. For plausible values of 2-3 kpc, the outflow rate would be $75 -110$~M$_{\odot}$~yr$^{-1}$. In comparison, the \FeII* emission has a characteristic size of $\sim 4$~kpc. 
Overall, the outflow rate is comparable to the star formation rate of 78~M$_{\odot}$~yr$^{-1}$.

\section{Conclusions}  
\label{sect:conclusions}

The direct detection of \ion{Fe}{II}* emission from the spatially-resolved MUSE HDFS galaxy ID\#13 at $z = 1.29$ opens a new avenue for studying galactic outflows in emission.
From an analysis of the deepest MUSE field so far (27 hours), the properties of this individual galaxy, including the inclination, stellar mass, star formation rate, and gas kinematics, are well characterized (Table~\ref{table:galProps}).
This galaxy has a low inclination ($i\sim$33 deg), \mstar, and \sfr.

Using the 1D integrated spectrum and 2D pseudo-narrow band images, we identify signatures of winds in emission from the \ion{Fe}{II}*, \ion{Fe}{II}, and \ion{Mg}{II} transitions and investigate the wind morphology and extent.   
Specifically, we find:
\renewcommand{\labelitemi}{$\bullet$}
\begin{itemize} \itemsep2pt
\item 
The star formation rate surface density from [\ion{O}{II}]\,$\lambda\lambda3727,3729$ is 
\sigSFR, well above the threshold for galactic winds \citep{2012ApJ...761...43N, 2014ApJ...784..108B}.
\item Asymmetric \ion{Fe}{II}, \ion{Mg}{II}, and \ion{Mg}{I} self-absorptions in the MUSE 1D spectrum have a strong blue wing that extends beyond $\sim -700$~km~s$^{-1}$. The profiles are well-fitted with three components at $-660$~km~s$^{-1}$,  $-350$~km~s$^{-1}$, and $+9$~km~s$^{-1}$ (Figure~\ref{absFits}). These blue-shifted absorptions indicate outflowing material along the line of sight, and we estimate a mass outflow rate in the range of $45 - 110$~M$_{\odot}$~yr$^{-1}$.
\item Emission infilling does not appear to be present because (i) all absorptions have very similar shapes, whereas infilling is varies significantly for different transitions \citep{2015ApJ...815...48Z} and (ii) the strongest component for all absorptions  (including \FeII\,$\lambda2600$ and $\lambda$2382) is close to the galaxy systemic redshift (Figure~\ref{absFits}).
A quantitative analysis following the \citet{2015ApJ...815...48Z} empirical method shows that emission infilling could impact the \FeII$\lambda$2600 (\FeII$\lambda2383$) rest-frame equivalent width by at most 10\%\ (25\%), respectively, and less for the other \FeII\ transitions.
\item Non-resonant \ion{Fe}{II}* emission from the $\lambda2365$, $\lambda2396$, $\lambda2612$, and $\lambda2626$ transitions have fluxes of $1.2 - 2.4 - 1.5 - 2.7 \times10^{18}$~\flux, respectively, and flux ratios that are consistent with optically thick gas \citep{2014ApJ...793...92T}.  The \ion{Fe}{II}*\,$\lambda2632$ transition has a 1-$\sigma$ flux limit of $< 8 \times \rm 10^{-19}$~\flux. Contrary to stacked spectra \citep[e.g.][]{2015ApJ...815...48Z}, the \FeIIs\ emission in this galaxy appear to be symmetric and well-centered on the galaxy  \OII\ systemic redshift (Figure~\ref{emissFits}).
\item After stacking the four non-resonant \ion{Fe}{II}* emission lines, we obtain the first spatially-resolved 2D map of this non-resonant emission from a $z\sim 1$ galaxy (Figure ~\ref{sbMaps}). The \ion{Fe}{II}* emission is more extended  than the stellar continuum or [\ion{O}{II}] emission. The \ion{Fe}{II}* emission half-light radius is $R_{1/2,\,\FeIIs}=$~\rhalfFeiikpc~kpc, about 50\%\ larger than that of the continuum which has $R_{1/2,\star}=$~\rhalfContkpc~kpc.
The \FeIIs\ emission has a different morphology; it is more elongated in the direction that roughly corresponds to the galaxy minor kinematic axis. 
\item 
The \ion{Fe}{II}* emission displays a velocity gradient along the kinematic minor axis, and the blue wing of the emission contours reveals a C-shape pattern in a $p-v$ diagram from a pseudo-slit extracted along this axis (Figure~\ref{pvPlots}).
These features are consistent with a conical outflow.
\item Comparing the observed emission and absorption properties with predictions
(Figure~\ref{eqWcomp}) from the radiative transfer models of \citet{2011ApJ...734...24P} suggests that the isotropic fiducial wind model fails, but that a biconical wind model including a dusty ISM component could more likely reproduce the observations from galaxy ID\#13.
\end{itemize}
This geometry agrees with a growing body of models and observations that suggest outflowing gas driven by supernovae explosions escapes the disk preferentially along the galaxy minor axis in a bi-conical flow \citep[e.g.,][]{2010AJ....140..445C, 2011ApJ...743...10B, 2012MNRAS.426..801B, 2013ApJ...774...50K, 2012ApJ...760L...7K, 2012ApJ...760..127M}.

\ion{Fe}{II}* emission from the MUSE HDFS galaxy ID\#13 was identified serendipitously, but by systematically searching through field galaxies in similar IFU data sets it will be possible to construct samples of $z \sim 1$ galaxies that each show evidence of outflows in emission. Observational constraints from these samples can then drive improvements to models of galactic-scale outflows.

\begin{acknowledgements}
      This work has been carried out thanks to the support of the ANR FOGHAR (ANR-13-BS05-0010-02), the OCEVU Labex (ANR-11-LABX-0060), and the A*MIDEX project (ANR-11-IDEX-0001-02) funded by the ``Investissements d'avenir'' French government program. 
      NB acknowledges support from a Career Integration Grant (CIG) (PCIG11-GA-2012-321702) within the 7th European Community Framework Program.
      RB acknowledges support from the ERC advanced grant 339659-MUSICOS.
      JB is supported by FCT through Investigador FCT contract IF/01654/2014/CP1215/CT0003, by Funda{\c{c}}{\~a}o para a Ci\^encia e a Tecnologia (FCT) through national funds (UID/FIS/04434/2013), and by FEDER through COMPETE2020 (POCI-01-0145-FEDER-007672).
      BE acknowledges support from the ``Programme National de Cosmologie and Galaxies'' (PNCG) of CNRS/INSU, France.
	  RAM acknowledges support by the Swiss National Science Foundation.
      JR acknowledges support from the ERC starting grant 336736-CALENDS.
      \end{acknowledgements}

%
%

\bibliographystyle{aa} 
\bibliography{FeII_Biblio} 

\end{document}